\documentclass[aps,pre,showpacs,twocolumn,superscriptaddress]{revtex4}

\usepackage{epsfig,amsmath,amssymb,epsf,bm,graphics}
\usepackage{mathrsfs,txfonts}
\usepackage{float}
\usepackage{epstopdf}

\begin{document}

\title{Extensions of Effective Medium Theory of\\ Transport in
Disordered Systems}
\author{V. M. Kenkre}
\affiliation{Consortium of the Americas for Interdisciplinary Science and
         Department of Physics and Astronomy, University of New Mexico, Albuquerque,
         New Mexico 87131, USA}
\author{Z. Kalay}
\affiliation{Consortium of the Americas for Interdisciplinary Science and
         Department of Physics and Astronomy, University of New Mexico, Albuquerque,
         New Mexico 87131, USA}
\author{ P. E. Parris}
\affiliation{Consortium of the Americas for Interdisciplinary Science and
         Department of Physics and Astronomy, University of New Mexico, Albuquerque,
         New Mexico 87131, USA}
\affiliation{Department of Physics, Missouri University of Science \& Technology, Rolla, MO 65409, USA}

\begin{abstract}
Effective medium theory of transport in disordered systems, whose basis is the replacement of spatial disorder by temporal memory, is extended in several practical directions. Restricting attention to a 1-dimensional system with bond disorder for specificity, 
a transformation procedure is developed to deduce, from given distribution functions characterizing the system disorder, explicit expressions for the memory functions. It is shown how to use the memory functions in the Lapace domain forms in which they first appear, and in the time domain forms which are obtained via numerical inversion algorithms, to address time evolution of the system beyond the asymptotic domain of large times normally treated. An analytic but approximate procedure is provided to obtain the memories, in addition to the inversion algorithm. Good agreement of effective medium theory predictions with numerically computed exact results is found for all time ranges for the distributions used except near the percolation limit as expected. The use of ensemble averages is studied for normal as well as correlation observables. The effect of size on effective medium theory is explored and it is shown that, even in the asymptotic limit,  finite size corrections develop to the well known harmonic mean prescription for finding the effective rate. A percolation threshold is shown to arise even in 1-d for finite (but not infinite) systems at a concentration of broken bonds related to the system size. Spatially long range transfer rates are shown to emerge naturally as a consequence of the replacement of spatial disorder by temporal memories, in spite of the fact that the original rates possess nearest neighbor character. Pausing time distributions in continuous time random walks corresponding to the effective medium memories are calculated.

\end{abstract}
\date{\today}
\maketitle

\section{Motivation for the Study}
Description of the movement of excitations and quasiparticles is crucial
to the study of a variety of disciplines in physics and allied
sciences \cite{dresden}. Conductivity in metals and semiconductors, energy transport in molecular
aggregates, atomic motion in ceramic materials, molecular hopping in
cell membranes, all present a diverse variety of contexts in which
such description is indispensable. At a sufficiently macroscopic
level, the description is often provided by a Master equation of the
type
\begin{equation}
\frac{dP_m(t)}{dt}=\sum_nF_{mn}P_n(t)-F_{nm}P_m(t)
\label{me}
\end{equation}
which governs the evolution of the probabilities $P_m(t)$ of
occupation of site $m$ by the moving entity at time $t$ via the
transition rates $F_{mn}$. Here $m$ is typically a vector index in the appropriately dimensioned space. The method of solution of such an equation
relies on the diagonalization of the so-called $A$-matrix defined
through $A_{mn}=-F_{mn}$ for $m\neq n$ and $A_{mm}=\sum_nF_{nm}$. One
can always formally write the solution of the probability vector $P(t)
$ from its initial value $P(0)$ as
\begin{equation}
P(t)=e^{-At}P(0).
\label{a}
\end{equation}
If the system is ordered (quasiparticle moving on a crystal lattice), this
solution becomes practical because the diagonalization can be
performed via discrete Fourier transforms and the $k$th mode of the
probabilities $P^k(t)=\sum_mP_m(t)e^{ikm}$ can be written as
\begin{equation}
P^k(t)=e^{-A_kt}P^k(0).
\end{equation}
Inversion into explicit $P_m(t)$'s is straightforward, and the
specifics of the dynamics of the $A$-matrix are evident through the
eigenvalues $A^k$ of the latter. An alternative way of analyzing the transport is via random walks. The relationship between a random walk description and a Master equation description was given long ago by Bedeaux et al. \cite{bedeaux}.

Physical systems in which disorder cannot be neglected are rampant in
nature. Whether the lack of order arises because some transfer bonds
are weaker or stronger than others, or whether the quasiparticles
must surmount barriers at some locations but not at others, disorder
must be seriously taken into account in the description of these
systems. A natural way is to replace the given system by a
corresponding ordered problem with temporal memory. What this means is that
the original time-local disordered problem, given by Eq. (\ref{me}), is
replaced by the so-called generalized master equation (GME),
\begin{equation}
\frac{dP_m(t)}{dt}=\int_0^t ds \sum_n\mathcal{W}_{mn}(t-s)P_n(s)-\mathcal{W}_{nm}(t-s)P_m(s),
\label{gme}
\end{equation}
where the memory functions $\mathcal{W}_{mn}$ are of the form $
\mathcal{W}_{m-n}$, i.e., translationally invariant,  the replacement
equation being, therefore, soluble via discrete Fourier transforms.
Then, in the  Laplace domain ($\epsilon$ is the Laplace variable and tildes denote Laplace transforms), the
counterpart of Eq. (3) is
\begin{equation}
\tilde{P}^k(\epsilon)=\frac{P^k(0)}{\epsilon+\tilde{\mathcal{A}}_k
(\epsilon)},
\end{equation}
where $\mathcal{A}_{mn}=-\mathcal{W}_{mn}$ for $m\neq n$, $\mathcal{A}_{mm}=\sum_n\mathcal{W}_{nm}$, and $\mathcal{A}_k$ is the Fourier transform of $\mathcal{A}_{m-n}$. Generalized master equations were introduced in the sixties to understand the derivation of the irreversible Master equation from reversible quantum mechanics and a comparative review of methods has been given by Zwanzig \cite{gme}. Needless to say, depending on one's choice, one could employ,  for the description of transport in disordered systems, 
continuous time random walk (CTRW) formalisms \cite{mw} instead of GME formalisms \cite{kms} in
equations such as (\ref{gme}). The two ways of description, CTRW's and GME's  have long been shown \cite{kk} to be entirely
equivalent to each other in fully arbitrary (rather than space-time decoupled) forms since 1974. The space-time decoupled case appeared in Kenkre et al \cite{kms} and the
demonstration of the equivalence for the general (arbitrarily coupled) form was given
in Kenkre and Knox \cite{kk} (see, for instance, their Eqs. (43)). It appears that the generalization in ref. \cite{kk} was missed by several different authors who reported it independently but subsequently \cite
{reiss,landman}, even six years later \cite{ks}. Some of this was commented on in brief in ref. \cite {K} (see Eqs. (27-31) of the latter reference). 

Two questions are important to answer in the context of this
program of the description of a disordered system. To
what extent is the replacement of the spatial disorder by temporal
memories possible and meaningful even in principle? And what is the
prescription to calculate the memories and effective transfer rates
given appropriate information about the disorder in the particular
system? Without the first, it is senseless to begin. Without the
second, the study is useless.

The first question can be answered quite trivially on a little
reflection. Consider Eq. (1) solved. By assigning the solutions for the probabilities, $P_m(t)$, to an appropriate \emph{ordered} lattice, carry out the direct Fourier transform to obtain $P^k(t)$. Put the Laplace transform of the latter into
\begin{equation}
\tilde{\mathcal{A}}_k=\frac{P^k(0)}{\tilde{P}^k(\epsilon)}-\epsilon
\label{prescr}
\end{equation}
and Fourier and Laplace invert to get the (translationally invariant) memories $\mathcal{A}_{mn}$, equivalently $\mathcal{W}_{mn}$ appearing in the GME (\ref{gme}). The presence of initial conditions in the above prescription means that each possible set of initial conditions would have a corresponding set of memory functions, a situation which is obviously unacceptable for practical purposes. However, in order to turn Eq. (\ref{prescr}) into a practical prescription for computing memories which is independent of initial conditions, all that is necessary is to carry out an ensemble average over the possible realizations of disorder, compatible with what is known (for instance a distribution function) about the disorder. Such an average makes the system  translationally invariant after the average. Then the first term in the right hand side of Eq. (\ref{prescr}) which is the reciprocal of the Fourier and Laplace transform of the (ensemble averaged) propagator, is independent of initial conditions. The propagators directly lead to the memories.

This is precisely the method devised long ago by Kenkre \cite{klongrange,kbook} to obtain exact expressions for memory functions analytically for a quantum mechanical (not disordered) system, although no ensemble average was involved in that context. Because equations (\ref{me}) and (\ref{gme}) as well as the operation of averaging over configurations are linear, it quite unnecessary to make any 
assumptions or offer demonstrations to be able to state with certainty that the replacement program is possible. Analyzing the problem from the viewpoint of the application of projection techniques \cite{zw} to the problem, it also becomes clear from Zwanzig's formal theory that a memory will automatically appear in a  closed description of \emph{any} quantity that is formally projected from another whose evolution equation is time-local. Here the projection is represented by an ensemble-average over disordered realizations. This too requires no calculation, only a moment's reflection. The initial condition problem, rarely discussed in the disorder context, also makes its appearance in the projection formalism \cite{zw,kinit}. It appears as a separate term. In the original context \cite{zw} it is removed through the initial random phase or diagonality assumption. In our present disorder context it disappears on carrying out the ensemble average we have mentioned above. 

What is really necessary in the sense of calculations comes to the second question we  have posed above, i.e., the finding of an \emph{explicit} and \emph{practical} prescription that would allow one to go from information about the disorder in the real system to the memories (or pausing time distribution functions) in the replacement problem. Very few instances of such a prescription exist in the literature, a noteworthy attempt  being in the early work of Scher and Lax \cite{sl} that gave
support to the well known  theory of Scher
and Montroll \cite{sm}: on a phenomenological basis, the latter addressed with great success unexplained
puzzles of transport in xerographic materials.  Known
information about transfer rates between randomly located sites is
converted  via an approximation scheme in the appendices of ref. \cite{sl}, into the
continuous time random walk pausing time description. That is the kind of prescription that one needs in developing a usable theory.

The present paper focuses on a different manner of converting disorder information into temporal memories that has to do with the venerable subject of effective medium theories (EMT) \cite{ancient,other,kirk,acreview,amherst,olax,parris,machta,zw2,kgran}. We provide the essential background on EMT in section 2, along with a prescription we provide in a particularly convenient form that transforms the disorder into explicit memory functions via a double Laplace transform procedure. Our prescription facilitates the extraction of the new results we present in subsequent sections.  The spirit of the investigations we present is most akin to, among  early attempts that have discussed memory functions in the EMT context \cite{ks,hkk}, the work of Haus and Kehr \cite{kehr,kehr2}.

\section{Explicit Disorder-to-Memory Transform}

Effective medium theories are unabashedly approximate, i.e., do not claim to provide an exact solution of the problem. They sacrifice exactness for practicability, i.e., prefer usefulness to avoidance of approximations. One of the first instances of their application is by Bruggeman \cite{ancient} but many later and independent reports may also be found \cite{other}. The basic idea, explained in many textbooks and reviews \cite{acreview}, is to assume that the memory represents an effective ordered medium in a mean field sense with a magnitude (of the memory) which is such that  any departures, introduced in keeping with whatever information is known about the disorder, average out to zero, thus establishing the ordered system as representing a variationally optimum limit. 

To understand this concretely, we consider from now on in this paper a  1-dimensional case of Eq. (\ref{me}) with bond disorder,
\begin{equation}
\frac{dP_m(t)}{dt}=F_{m+1}[P_{m+1}(t)-P_m(t)]+F_{m}[P_{m-1}(t)-P_m(t)],
\label{1d}
\end{equation}
the disorder being expressed via a distribution $\rho(f)$. What this means is that any transition rate $F_m$ can have any positive value $f$ with probability density $\rho(f)$ normalized such that $\int_0^{\infty}\rho(f)df=1$. No correlations exist in the actualization of rates at different locations. The replacement of the disordered time-local system by an ordered system with memory then proceeds by writing in place of Eq. (\ref{1d}), a GME
\begin{equation}
\frac{dP_m(t)}{dt}=\int_0^t ds \mathcal{F}(t-s)[P_{m+1}(s)+P_{m-1}(s)-2P_m(s)],
\label{gme2}
\end{equation}
which is translationally invariant and describes elemental transfer interactions that are nearest neighbor as in the original (disordered) problem.

When applied to the present system, the general effective medium concept requires the following procedure. One  evaluates the probability propagators  for two different systems: the ordered system obeying Eq. (\ref{gme2}), and a system obeying Eq. (\ref{gme2}) augmented by terms that represent a single disordered bond. Transport across that bond  occurs not through the memory $\mathcal{F}(t)$ but through a rate $f$ drawn from the distribution $\rho(f)$. One averages the latter propagators over the distribution, i.e., carries out an integration of the result with weight $\rho(f)$, and demands that the average equal the corresponding ordered propagators. Details may be found in the references given and lead straightforwardly to
\begin{equation}
\int_0^{\infty}df\frac{\rho(f)}{1+2[f-\tilde{\mathcal{F}}(\epsilon)][\tilde
{\psi}_0(\epsilon)-\tilde{\psi}_1(\epsilon)]} =1.
\label{everyone}
\end{equation}

The above equation is an implicit equation for the memory $\mathcal{F}$ that can be obtained in principle from the given probability distribution function $\rho(f)$, the quantities $\psi_0$ and $\psi_1$ being the propagators of the ordered system: the probability of remaining on the site initially occupied is $\psi_0$ whereas the probability of occupation of the adjacent site is $\psi_1.$ Equation (\ref{everyone}) is the same as Eq. (22) of ref. \cite{acreview}, or Eq. (5.4) of ref. \cite{kirk}, or Eq. (7) of ref. \cite{hkk}, or Eq. (3.17) of ref. \cite{olax} or Eq. (38) of ref. \cite{amherst}. Through a simple manipulation we  rewrite it first as
\begin{equation}
\int_0^{\infty}df\frac{\rho(f)}{f+\tilde{\mathcal{F}}(\epsilon)\left[\frac
{\epsilon\tilde{\psi}_0(\epsilon)}{1-\epsilon\tilde{\psi}_0(\epsilon)}
\right]} =\frac{1}{\tilde{\mathcal{F}}(\epsilon)}\left[1-\epsilon\tilde{\psi}_0
(\epsilon)\right]
\end{equation}
and then, by introducing a quantity $\xi$, in the remarkably simple and convenient form
\begin{equation}
\int_0^{\infty}df\,\,\frac{\rho(f)}{f+\xi}=\frac{1}{\tilde{\mathcal{F}}+\xi}.
\label{theeqn}
\end{equation}

The quantity $\xi(\epsilon,\tilde{\mathcal{F}})$ is a function of both $\epsilon$ and of $\tilde{\mathcal{F}}(\epsilon)$ since the selfpropagator $\tilde{\psi}_0$ depends explicitly on $\tilde{\mathcal{F}}(\epsilon)$ as well as on $\epsilon$. Generally,
\begin{equation}
\xi=\tilde{\mathcal{F}}(\epsilon)\left[\frac
{\epsilon\tilde{\psi}_0(\epsilon)}{1-\epsilon\tilde{\psi}_0(\epsilon)}
\right].
\label{genxi}
\end{equation}
For the infinite 1-d chain with nearest neighbor rates, given that, in this case, $\epsilon\tilde{\psi}_0$ equals  $[1+4\tilde{\mathcal{F}}(\epsilon)/\epsilon]^{-1/2}$, one has the specific expression
\begin{equation}
\xi=\frac{\epsilon}{4}\left(1+\sqrt{1+\frac{4\tilde{\mathcal{F}}(\epsilon)}{\epsilon}}\right).
\label{specxi}
\end{equation}

As we will see below, this restatement (\ref{theeqn}) of the basic EMT equation (\ref{everyone}) allows us to obtain a number of our results in a straightforward fashion.  
With very few exceptions in the literature, the result (\ref{everyone}) is used in the long-time limit and therefore involves the Markoffian replacement of $\mathcal{F}(t)$ by $\delta(t)[\int_0^{\infty}\mathcal{F}(s)ds]$. This is equivalent to the $\epsilon \rightarrow 0$ limit. By an Abelian theorem $\epsilon\tilde{\psi}_0$ becomes identical to the $t \rightarrow \infty$ limit of the selfpropagator which is zero if the system considered is an infinite chain. Equation (\ref{theeqn}) then reduces to the well known effective medium theory result \cite{zw2} that the effective transfer rate $F_{eff}=\int_0^{\infty}\mathcal{F}(s)ds=\tilde{\mathcal{F}}(\epsilon \rightarrow 0)$ equals the harmonic mean of the disordered $f$'s:
\begin{equation}
\frac{1}{F_{eff}}=\frac{1}{\tilde{\mathcal{F}}(0)}=\int_0^{\infty}df\,\,\frac{\rho(f)}{f} 
\label{harmo}.
\end{equation}
By contrast, our interest in the present paper is to extract new information from the memory equation \emph{without taking the Markoffian limit}, and to go beyond common uses of
effective medium theory. We will derive some general features of the EMT memory in section 3, describe our extensions of the theory for times that are not asymptotic in section 4, analyze the emergence of spatially long range memories in section 5, study finite size effects in section 6, and present conclusions  in section 7.

The first of the results of our present investigation is the reformulation implicit in Eq. (\ref{theeqn}) interpreted as a \emph{transform} of the distribution function $\rho(f)$ (disorder information) into the effective medium quantity $\mathcal{F}(t)$ (temporal memory). Specifically, we can regard Eq. (\ref{theeqn}) as related to a double Laplace transform. One applies the \emph{direct} Laplace transform twice: first to $\rho(f)$, with a dummy variable $y$ as the Laplace variable, to obtain $g(y)$, and then to $g(y)$ with  $\xi$ as the Laplace variable to obtain $h(\xi)$:
$$
g(y)=\int_0^{\infty} \rho(f)e^{-yf}df;\,\,\,\,\,\,\,\,h(\xi)=\int_0^{\infty} g(y)e^{-\xi y}dy.
$$

The prescription for extracting the memory $\mathcal{F}(t)$  in the EMT equation (\ref{gme2}) from the disorder distribution $\rho(f)$ consists, thus, of computing the double transform $h(\xi)$ of the disorder distribution, equivalently performing the integral on the left side of (\ref{theeqn}), and inverting into the time domain  the memory transform $\tilde{\mathcal{F}}(\epsilon)$ after solving for it from the implicit equation
\begin{equation}
h(\xi)=\frac{1}{\tilde{\mathcal{F}}(\epsilon)+\xi(\epsilon,\tilde{\mathcal{F}})}.
\end{equation}
One has, thus, a practical prescription to obtain the $t$ dependence of the memory $\mathcal{F}(t)$ from the disorder  $\rho(f)$.

The usefulness of the form of the basic equation we have presented, Eq. (\ref{theeqn}), should be already clear by comparison to the well-known asymptotic result for the effective rate Eq. (\ref{harmo}) and noticing from Eq. (\ref{genxi}) or Eq. (\ref{specxi}) that in the asymptotic limit $\xi$ vanishes. Further uses are reported in the rest of the paper.

\section{Nature of the Effective Medium Memory Functions}
We have applied the prescription of Eq. (\ref{theeqn}) to various distribution functions $\rho(f)$ to obtain $\widetilde{\mathcal{F}}(\epsilon)$ and discovered that the results share a number of common features. These features become apparent on inverting the transform to obtain  $\mathcal{F}(t),$ the memory function in the time domain, and can be understood as we show below. The numerical inversion scheme we use is \cite{gaver1966,stehfest1970,abate_whitt2006}
\begin{eqnarray}
\mathcal{F}_{g}(t,M)&=&\frac{\ln 2}{t} \sum_{k=1}^{2M}\zeta_k \widetilde{\mathcal{F}}\left( k\frac{\ln 2}{t} \right), \nonumber \\
\zeta_{k}&=&(-1)^{M+k}\sum_{j=\lfloor (k+1)/2 \rfloor}^{\min(k,M)}\frac{j^{M+1}}{M!}\binom{M}{j}\binom{2j}{j}\binom{j}{k-j} 
\label{gaver_stehfest}. 
\end{eqnarray}
where $\mathcal{F}_{g}(t,M)$ is the approximation to the Laplace inverse of $\widetilde{\mathcal{F}}(\epsilon)$. The precision required to sum the series, i.e. the number of significant digits, is $2.2M$ while the precision of the resulting expression is $0.90M$. So, if one uses double precision in the calculations, the value of $M$ should not be larger than $7$. For a detailed discussion on the numerical inversion of  Laplace transforms, see ref. \cite{abate_whitt2006}. Note that the function is only evaluated at the real and positive values of the Laplace variable $\epsilon$. 

The result of the Laplace inversion is always that the EMT memory $\mathcal{F}(t)$  consists of two pieces, a $\delta$-function at the origin of time ($t=0$) and a part that is negative but finite. A schematic depiction is given in Fig. 1. In order to understand this and other qualitative, and some quantitative, aspects of $\mathcal{F}(t)$ from general arguments, consider the actual system evolution equation (\ref{1d}) on the one hand, and the representative EMT equation (\ref{gme2}) on the other, both for an initial occupation of only the site $m$. Let us first evaluate the first time derivative of $P_m(t)$  at the initial time. The respective results are
\begin{equation}
\left[\frac{dP_m(t)}{dt}\right]_{t=0}=-(F_m+F_{m+1})
\label{1}
\end{equation}
for the actual Master equation, and
\begin{equation}
\left[\frac{dP_m(t)}{dt}\right]_{t=0}=\int_{0-}^{0+} ds \mathcal{F}(t-s)[P_{m+1}(s)+P_{m-1}(s)-2P_m(s)]
\label{2}
\end{equation}
for the representative EMT equation. A configuration average over the distribution $\rho(f)$ converts the right hand side of Eq. (\ref{1}) into $-2\langle f \rangle=-2\int df\rho(f)f$. It is impossible for Eq. (\ref{2}) to yield a non-zero result (because of the limits of integration) unless $\mathcal{F}(t)$ contains a $\delta$-function at the origin. As  Eqs. (\ref{1}) and (\ref{2})  must yield results that equal each other, we deduce that the form of the  EMT memory function is
\begin{equation}
\mathcal{F}(t)=\langle f \rangle\delta(t)-Q(t).
\label{general}
\end{equation}

\begin{figure}[]
\centering
\includegraphics{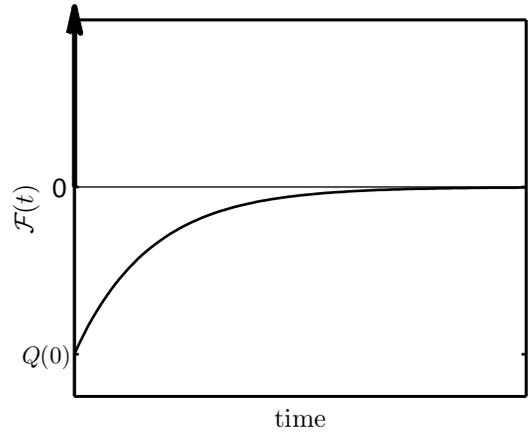}
\caption{Shape of the effective medium memory function $\mathcal{F}(t)$ showing the delta function  of strength $\langle f \rangle$ at the origin and the negative piece $Q(t).$ The time integral from $0$ to $\infty$ of the memory $\mathcal{F}(t)$ is $1/\langle 1/f \rangle$.}
\label{genmemory}
\end{figure}

The origin of the $\delta$-function at $t=0$ is clear from the above analysis. That the additional part must have a time integral for all time which is negative follows from the general result (\ref{harmo})  that the integral over all time of $\mathcal{F}(t)$ is the harmonic mean $1/\langle 1/f \rangle=\int df \rho(f)/f$ which is always smaller \cite{feetnote} than the arithmetic mean $\langle f \rangle$. 

We also note that the integral of $Q(t)$ over all time is now determined:
\begin{equation}
\int_0^{\infty}Q(t)dt=\langle f \rangle-\frac{1}{\langle 1/f \rangle}.
\end{equation}
Additional information can be obtained in this exact manner about the memory function, for instance, the initial value of $Q(t)$. Differentiation of Eq. (\ref{1d}) with respect to time yields the initial \emph{second} time derivative
\begin{equation}
\left[\frac{d^2P_m(t)}{dt^2}\right]_{t=0}=2\left(F^2_{m+1}+F^2_{m}+F_{m+1}F_{m}\right).
\end{equation}
Similarly, differentiation of  the EMT generalized master equation (\ref{gme2}) yields, after a configuration average,
\begin{equation}
\left[\frac{d^2P_m(t)}{dt^2}\right]_{t=0}=6\langle f \rangle^2+2Q(0).
\end{equation}
Carrying out the configuration average of the former result, which gives $4\langle f^2 \rangle+2\langle f \rangle^2,$ and equating the two values of the second time derivatives at the initial time, we can evaluate  $Q(0)$ \emph{exactly} for any distribution function as
\begin{eqnarray}
Q(0)&=&2\left[\langle f^2 \rangle-\langle f \rangle^2\right] \nonumber \\
&=&2\left[\left(\int df\rho(f)f^2\right)-\left(\int df\rho(f)f\right)^2\right].
\end{eqnarray}
It is also straightforward to continue in this manner with further differentiations to obtain exact initial values of higher derivatives of $Q(t)$. For instance, in terms of the $A$-matrix appearing in Eq. (\ref{a}), we can evaluate the initial value of the $r$th derivative of $P_m$ via
$$ \left[\frac{d^rP_m}{dt^r}\right]_0=(-1)^r(A^r)_{mm}$$
and proceed as shown above with configuration averages. For our nearest neighbor rate system we have $A_{mn}=-F_{m+1}\delta_{m,n+1}-F_m\delta_{m,n-1}+(F_{m+1}+F_m)\delta_{m,n}$. 

We do not carry out this program here but use the limited information gathered above to develop a simple analytical approximation to the memory cast in the form of a difference of a term proportional to a delta function and another to an exponential. This `exponential' approximation to the EMT memory for any given distribution of the rates is
\begin{equation}
\mathcal{F}_a(t)=\langle f \rangle\delta(t)-2\left[\langle f^2 \rangle-\langle f \rangle^2\right]e^{-t\left(\frac{2(\langle f^2 \rangle-\langle f \rangle^2)}{\langle f \rangle-(\langle 1/f \rangle)^{-1}}\right)}.
\label{expapprox}
\end{equation}
The subscript $a$ clarifies that the memory is approximate. While the precise shape of the actual memory function will not be captured by our approximation (\ref{expapprox}), examples to be given in the next section will make clear that the approximate memory can be remarkably good.

The general behavior of the time dependence of the memory function consisting of a decay (infinitely fast for our system) to negative values and then a rise which is often relatively slower is typical in many systems. It is usually encountered in studies of the velocity autocorrelation $\langle v(t)v \rangle$ which is, needless to say, very closely related (in our case simply proportional) to the memory function. The small time behavior represents initial transfer at a higher rate; the subsequent behavior is affected by disorder or imperfections in the system as they are encountered in the motion. Indeed, the  velocity autocorrelation for a random walker completely confined to a finite space exhibits this very behavior, the overall integral of $\langle v(t)v \rangle$ for all time being precisely zero because of the confinement: the mean square displacement saturates in this case (see, e.g., ref. \cite{sheltrawk} for a nuclear magnetic resonance context.)

\section{Beyond Asymptotic Description}
Research reported on the basis of effective medium theory
is almost entirely focused on an asymptotic description of quantities
such as the diffusion constant. There have been a few exceptions as in the work with focus on the ac conductivity,  e.g., by Odagaki and Lax \cite{olax}, and others \cite{acreview}, the publications of  Haus and Kehr \cite{hkk,kehr,kehr2},
the anisotropic disordered systems studied by Parris \cite{parris}, and the
granular material stress work of Kenkre \cite{kgran}. One of the questions we
address in the present paper is how well effective medium theory works
for times \emph{outside of} the long and short time asymptotic domains, i.e.,
for all times. To address this problem we now compute the memories explicitly for three different disorder distribution functions $\rho(f)$, use those memories in the GME to calculate observables and compare the results to numerically obtained exact solutions of the disordered Master equation.

\subsection{Some Specific Distributions $\rho(f)$}
A natural distribution to consider is the multi-delta distribution
$$ \rho(f) = \sum_{i=1}^M \alpha_i \delta(f-f_i)
$$
wherein the nearest-neighbor transition rates  may take one of $M$ values $f_i$ each with a weight $\alpha_i$, with  $\sum_{i=1}^M \alpha_i=1$. We will focus particularly on the case $M=2$, so that 
\begin{equation}
\rho(f)=\alpha \delta(f-f_1) + (1-\alpha)\delta(f-f_2). \label{doubledelta}
\end{equation}
The arithmetic and harmonic means of the rates, additionally the mean of the square of the rates,  are given by
\begin{eqnarray}
\langle f \rangle&=&\alpha f_1  + (1-\alpha) f_2, \nonumber \\
\frac{1}{\langle 1/f \rangle}&=&\frac{f_1f_2}{\alpha f_2+(1-\alpha)f_1}, \nonumber \\
\langle f^2 \rangle&=&\alpha f_1^2  + (1-\alpha) f_2^2.
\label{meansdd}
\end{eqnarray}
The distribution $\rho(f)$ for this case is shown as the two arrows in Fig. 2.

The second particular distribution we consider is the gamma distribution (related closely to the Poisson distribution):
\begin{equation}
\rho(f)=\frac{\gamma^{n+1}}{\Gamma(n+1)}f^n e^{-\gamma f}  \label{gamma}.
\end{equation}
The arithmetic and harmonic means of the rates, and the mean of the square are 
\begin{eqnarray}
\langle f \rangle&=&\frac{n+1}{\gamma},  \\
\frac{1}{\langle 1/f \rangle}&=&\frac{n}{\gamma}, \\
\langle f^2 \rangle&=&\frac{(n+1)(n+2)}{\gamma^2}.
\label{meansgamma}
\end{eqnarray}
A plot of $\rho(f)$ itself is displayed for the particular case of $n=1$ and $\gamma=4$ in Fig. 2. 
\begin{figure}[]
\centering
\includegraphics{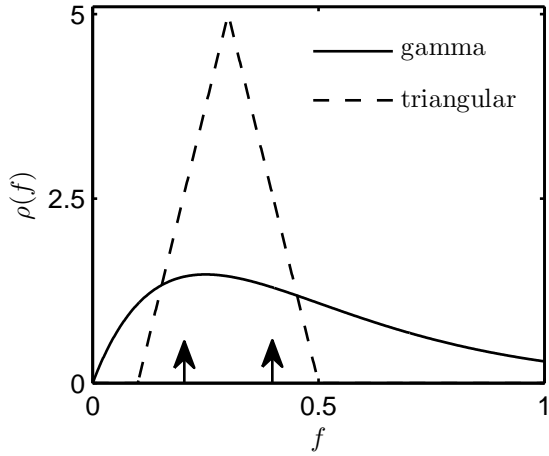}
\caption{Examples of probability distributions $\rho(f)$. The two arrows represent the double-delta distribution with equal weight $\alpha=1-\alpha=0.5.$ A gamma distribution with $n=1$ and $\gamma =4$ and a triangular distribution with $f_{0}=0.3$ and $b=0.2$ are depicted by the solid and dashed lines respectively. Units of $f$ in the plot are arbitrary and the same as those of $f_0$ and $b$ for the triangular distribution, and reciprocal to those of $\gamma$ for the gamma distribution.}
\label{distributions}
\end{figure}

The third case we consider is the triangular distribution given by
\begin{equation}
\rho(f)=\begin{cases} 
(f-f_0+f_b)/f_b^2& \text{$f_0-f_b \leq f \leq f_0 $}, \\
(-f+f_0+f_b)/f_b^2& \text{$f_0 \leq f \leq f_0+f_b$},  \\
0& \text{elsewhere}.  
\end{cases} \label{triangular}
\end{equation}
The minimum possible rate is $f_0-f_b$ and the maximum possible rate is $f_0+f_b.$ The distribution rises linearly from the minimum value with slope $1/f_b^2$ until it attains the value $1/f_b$ at $f=f_0$ and then descends with the same magnitude of the slope down to the maximum value. The meaning of $f_0$ is that it is the value of $f$ at the apex (and hence the mean of the distribution), and $f_b$ is half the length of the base of the triangle. The distribution is shown in Fig. 2 for $f_0=0.3$ and $f_b=0.2$. It leads to 
\begin{eqnarray}
\langle f \rangle&=&f_0,  \\
\langle f^2 \rangle&=&f_0^2+\frac{f_b^2}{6}, \\
\frac{1}{\langle 1/f \rangle}&=&\frac{f_b}{\ln\left( 1 + \frac{2f_b}{f_0-f_b} \right) + \frac{f_0}{f_b} \ln \left( 1 - \frac{f_b^2}{f_0^2} \right)}.
\label{meanstriangular}
\end{eqnarray}

\subsection{Evaluation of Memory Functions for Specific Cases of $\rho(f)$}
The approximation to the memory given by the formula (\ref{expapprox}) is easily evaluated for the three distributions by substituting in the formula the respective values of $\langle f \rangle$, $\langle f^2 \rangle$ and $1/\langle 1/f \rangle$. As one example, note that for the gamma distribution it is given by
\begin{eqnarray}
\mathcal{F}_a(t)&=&\left(\frac{n+1}{\gamma}\right)\delta(t)-2\left(\frac{n+1}{\gamma^2}\right) e^{-\frac{2(n+1)t}{\gamma}}.
\label{expmemgamma}
\end{eqnarray}

The memory function $\tilde{\mathcal{F}}(t)$, whether derived from (\ref{theeqn}) or the simpler (\ref{expapprox}), can be used immediately to calculate other, more directly observable, quantities. A useful quantity is the (dimensionless) mean square displacement $\langle m^2 \rangle=\sum_m m^2P_m(t)$ for initial localization at the origin. It is simply twice the double time integral of the memory:
$$ \langle m^2 \rangle=\sum_m m^2P_m(t)=2\int_0^t ds \int_0^s \mathcal{F}(y) dy.$$ The time-dependent diffusion coefficient $D(t)$, a quantity often used in transport theory to describe the instantaneous state of motion, may be defined as one half the product of the square of the intersite distance $a$ and the time derivative of the mean square displacement. It is proportional to a single time integral of the memory function:
$$ D(t)=\frac{a^2}{2}\left(\frac{d\langle m^2 \rangle}{dt}\right)=a^2\int_0^s \mathcal{F}(s) ds.$$
These have exact expressions in terms of $\langle f \rangle$ and $Q(t)$ appearing in Eq. (\ref{general}). If we use our simple exponential approximation for $Q(t)$, they become
\begin{eqnarray}
\label{msd}
\frac{\langle m^2 \rangle}{2}&=&\frac{t}{\left\langle1/f \right\rangle}+\frac{\left(\left\langle f \right\rangle -\left\langle1/f \right\rangle^{-1} \right)^2}{2(\langle f^2 \rangle-\langle f \rangle^2)}\left(1-e^{-t/\tau}\right),  \\
 \frac{D(t)}{a^2}&=&\left\langle f \right\rangle-\left(\left\langle f \right\rangle -\left\langle1/f \right\rangle^{-1} \right)\left(1-e^{-t/\tau}\right),
 \label{expmsdD}
\end{eqnarray}
where the time constant $\tau$ is given by $$ \tau=\left(\frac{\langle f \rangle-(\langle 1/f \rangle)^{-1}}{2(\langle f^2 \rangle-\langle f \rangle^2)}\right).$$
It is straightforward to get expressions particular to the distribution functions chosen. As expected, the mean square displacement starts out linearly with slope twice the arithmetic mean of the rates and ends up also linearly with slope twice the harmonic mean of the rates. Correspondingly, the time dependent diffusion constant decays from a higher to a lower value. 

There are a number of ways the above simple analysis can be put to use to extract physical information. For instance,  the mean square displacement of a walker  initially localized at a single site will first grow linearly but then saturate to a finite value at long times if there are broken bonds in the 1-d infinite system. Broken bonds correspond to a $\rho(f)$ that has a non-zero value at $f=0$ which means that there are bonds at which the transition rate is zero. In such a case, $1/\langle 1/f \rangle$, the harmonic mean of the rates, and consequently the long time $D(t)$, vanish. Equation (\ref{msd}) can then be used to extract the value at which the mean square displacement saturates at long times:
\begin{equation}
\lim_{t \rightarrow \infty} \langle m^2 \rangle=\frac{\left\langle f \right\rangle ^2}{\langle f^2 \rangle-\langle f \rangle^2}.
\label{parti}
\end{equation}
This consequence of the exponential approximation (\ref{expapprox}) to the memory is simply a case of the general EMT result
\begin{equation}
\lim_{t \rightarrow \infty} \langle m^2 \rangle=-2\lim_{\epsilon \rightarrow 0}\frac{d\tilde{Q}(\epsilon)}{d \epsilon}.
\label{satmsd}
\end{equation}
This may be proved from the Laplace transform of the non-delta part of the memory in Eq. (\ref{general}) via a Taylor expansion:
$$\tilde{\mathcal{F}}(\epsilon)=\langle f \rangle-\tilde{Q}(\epsilon)=\frac{1}{\langle 1/f \rangle}-\epsilon \left[\frac{d\tilde{Q}(\epsilon)}{d \epsilon}\right]_{\epsilon=0}-\frac{\epsilon ^2}{2} \left[\frac{d^2\tilde{Q}(\epsilon)}{d \epsilon ^2}\right]_{\epsilon=0}...$$
In the presence of broken bonds in 1-d, the harmonic mean of $f$'s vanishes. Since the mean square displacement is twice the double time integral of $\mathcal{F}(t)$, the limit $\epsilon \rightarrow 0$ and the use of an Abelian theorem establish Eq. (\ref{satmsd}) quite generally. If $\mathcal{F}(t)$ is expressed via the exponential approximation (\ref{expapprox}), the general result reduces to Eq. (\ref{parti}).

\begin{figure}[]
\centering
\includegraphics[width=3in]{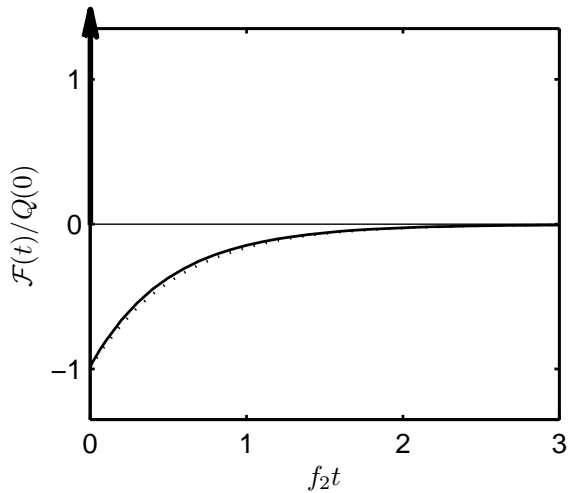}
\caption{Comparison of the exact (numerical) memory function with the approximate EMT given by our formula (\ref{expapprox}) for a double-delta distribution in which the concentration of the smaller of the rates is 0.9 and the ratio of the transition rates is 10. Plotted are the exact memory function (solid line) and the exponential approximation (dotted line). It is surprising how close the agreement is, given the coarse nature of the approximation.}
\label{goodexpapp}
\end{figure}

Despite what appears as an impressive agreement of the exponential approximation
that we see displayed in Fig. \ref{goodexpapp} for a double delta
distribution with $\alpha = 0.9$, and $f_{2}/f_{1}=10$, the approximation
generally will not capture the actual decay in time for all distribution
functions and  may be regarded only as a highly simplified manner of description.  For greater accuracy than can be provided by the relatively coarse approximation of Eq. (\ref{expapprox}), it is necessary to return to the prescription of Eq. (\ref{theeqn}), calculate $\tilde{\mathcal{F}}(\epsilon)$ through the solution of the implicit equation, and then invert the transform to obtain the memory. When the EMT memory is calculated in the Laplace domain via our prescription based on Eq. (\ref{theeqn}), the derived quantities $D(t)$ and $\langle m^2 \rangle$ can be obtained very simply in the Laplace domain by dividing $\tilde{\mathcal{F}}(\epsilon)$ by $\epsilon$ and $\epsilon ^2$ (except for proportionality constants) respectively.

The calculation of $\tilde{\mathcal{F}}$ from Eq. (\ref{theeqn}) is easy and analytically doable for the double-delta distribution. Defining the quantity $$\eta = (1-\alpha) f_1 + \alpha f_2,$$ we get the soluble cubic
\begin{gather}
\widetilde{\mathcal{F}}^3 - 2\widetilde{\mathcal{F}}^2\left( 2\eta^2/\epsilon +f_1 + f_2) \right) \nonumber \\ + \widetilde{\mathcal{F}}\left( 8\eta f_1 f_2 / \epsilon + \left( 2 f_1 f_2 +(f_1 + f_2)^2 - \eta^2 \right) \right) \nonumber \\ - 4f_1^2 f_2^2 / \epsilon - \left( 2f_1 f_2 (f_1 + f_2) - 2\eta f_1 f_2 \right) = 0. \label{double_delta_cubic} \end{gather}
Standard  analytic formulae lead to the appropriate solution which can then be numerically Laplace-inverted. We have carried out such a procedure in the next section. 

Similar procedures can be used for the gamma distribution and the triangular distribution. Explicit polynomials do not result for $\widetilde{\mathcal{F}}$ in those cases but the equations can be solved numerically and inverted. We have carried out these procedures for these two distributions as well and report the results after inversions into the time domain using Eq. (\ref{gaver_stehfest}).

\begin{figure}[]
\centering
\includegraphics{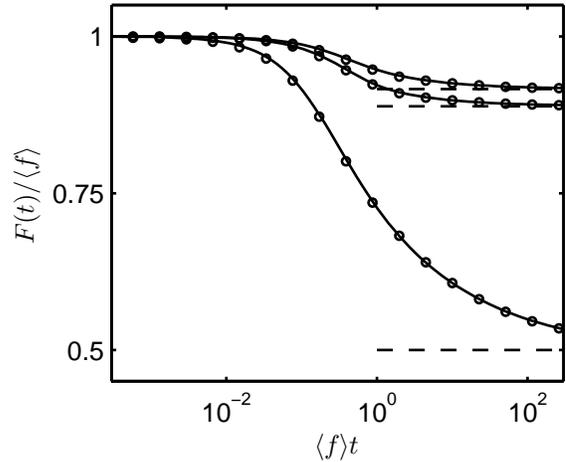}
\caption{Comparison of EMT predictions with exact results for three disorder distributions $\rho(f)$. Plotted is the time-dependent diffusion coefficient (normalized to its initial value) as a function of the dimensionless time $\tau=\left\langle f \right\rangle t$ for three cases of  $\rho(f)$ and a localized initial condition. Solid lines are the effective medium results and open circles correspond to the numerically exact  results obtained by averaging over 20000 different realizations of the disordered chain. Dashed lines on the right show the asymptotic values of the diffusion coefficient. From top to bottom, $\rho(f)$ is the double-delta distribution with $f_1/f_2=0.5$ and $\alpha=0.5$, a triangular distribtution with $f_0=0.3$, $f_b=0.2$, and a gamma distribution with $n=1$.The agreement is striking for all three cases, not only for extreme limits but for intermediate times as well. }
\label{figure1}
\end{figure}

\subsection{Comparison of EMT and Exact Solutions for All Times}
We now display the results of the predictions of effective medium theory and the numerically obtained exact evolution not only for long times as is usually done, but for short and intermediate times as well.  For each distribution we calculate the exact and the full EMT results.  The exact results are obtained via numerical matrix operations as explained elsewhere in this paper. With the exception of single-run studies to be reported further below, we repeat the operations tens of thousands of times, each time using a different realization of the chain. Then we  average over all runs to produce the quantity we desire. The effective medium theory prediction for that quantity is also determined via the effective medium memory function both in its full form as given from our Eq. (\ref{theeqn}), and from our analytic approximation, Eq. (\ref{expapprox}). 

We first treat the case when a single site is initially fully occupied. 

\subsubsection{Localized Initial Condition}
\begin{figure}[]
\centering
\includegraphics{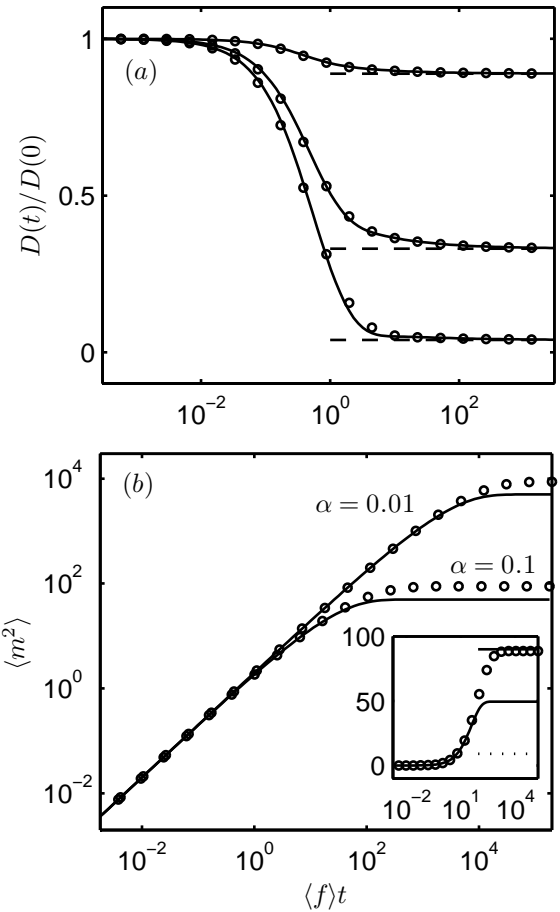}
\caption{Worst-case scenario comparison of EMT and exact results. In (a) we plot the time-dependent diffusion coefficient (normalized to its initial value) as a function of the dimensionless time $\tau=\left\langle f \right\rangle t$ for the double-delta distribution function  for $\alpha=0.5$ and, from top to bottom, $f_1/f_2=0.5$, $f_1/f_2=0.1$, $f_1/f_2=0.01$.  While good, the agreement gets worse for disparate $f$'s. To explore a regime in which the agreement is bad, in (b)  we consider two broken-bond systems (the ratio of the $f$'s being zero and therefore extreme) with two different concentrations $\alpha=0.1,0.01$ as shown. Plotted is the mean square displacement showing saturation at long times. Here $f_1=0$ and $f_2=0.2.$ Open circles correspond to the exact (numerical) solution obtained by averaging over 20000 different realizations of the disordered chain which consists of 801 sites. Solid lines are theoretical results from the EMT. The inset shows the  $\alpha=0.01$ case, the ordinate being on a linear scale in units of $10^4$. See text for discussion.}
\label{figure2}
\end{figure}

Figs. \ref{figure1} and \ref{figure2} display the comparison graphically, the quantity selected being the time-dependent diffusion coefficient normalized to its initial value: $D(t)/D(0)$. All three distributions are considered in Fig. \ref{figure1}. The agreement of the effective medium theory (solid lines) with the exact evolution (open circles) is remarkably good for all cases considered and for all intermediate times as well. The description appears thus excellent for the parameters considered for times that need not be asymptotic. 

In order to explore parameter values for which the agreement may \emph{not} be as good, we restrict ourselves to the double-delta distribution in (a) of Fig. \ref{figure2}, take the two possible rates $f_1$ and $f_2$ to occur with equal weight ($\alpha=0.5$),  but vary the ratio: $f_1/f_2=0.5,0.1,0.01$ as we go down the graph. EMT is still found to provide a fine description for all times but deviates more as the rates become more disparate. To drive this situation to an extreme where the EMT would serve worst, we consider a broken bond system in (b) of Fig. \ref{figure2}. This means we take $f_1=0$ and $f_2\neq0$ for different values of the concentration $\alpha$.  The large time value of $D(t)$ is now zero and the mean square displacement $\langle m^2 \rangle$ (proportional to the integral of $D(t)$) saturates. Physically, the saturation value measures the size of clusters (separated by broken bonds from other clusters) on which the walker is localized at long times. We have already obtained analytic expressions for the saturation value from the full EMT (in Eq. (\ref{satmsd})) and from the exponential approximation (in Eq. (\ref{parti})). Fig. 5b is an attempt at looking at EMT in the worst possible light by comparing the time evolution of $\langle m^2 \rangle$ predicted by it to that given by exact calculations. We do this for the broken bond case ($f_1=0$) for two concentrations $\alpha$ of broken bonds: $0.01$ and $0.1$ as shown. The main display in Fig. 5b shows the two $\langle m^2 \rangle$ curves. To make the discrepancy of the saturation value particularly clear, we show the inset in which the one case $\alpha=0.01$ is displayed on a semilogarithmic scale. The abscissa is the dimensionless time $\langle f \rangle t$ as in the main figure. The ordinate is $\langle m^2 \rangle$ on a linear scale, the values displayed as $0.5$ and $1$ being $5000$ and $10000$ (i.e., in units of $10^4$). The accumulated values of the mean square displacement, the localization cluster sizes, are $8.71 x10^3$ from the exact calculations but only  $5.00 x 10^3$ from the EMT: both are denoted by solid lines in the inset. The corresponding values for the $\alpha=0.1$ case are $88.6$ and $49.5$ respectively. The exponential approximation to the EMT is way off as it predicts $99$ for the $\alpha=0.01$ case and $9$ for  $\alpha=0.1$. (The latter is denoted by a dotted line in the inset.) This is to be expected from the crudeness of that approximation.

\subsubsection{Extended Initial Conditions for Single Runs}
An actual experiment in a real physical situation is performed not on an ensemble but on an individual system. How can EMT, which has at its root an ensemble average, provide a valid description for the experiment? Standard Gibbs-Boltzmann arguments do not help as an answer here because our interest in using EMT is not only for asymptotic times when the system might have completed the mixing process but for all times. One possible answer to this question might lie in the nature of the \emph{initial} condition. If it is \emph{extended in space}, various configurations of transition rates in a random system may be realized even at short times. With this idea in mind we now describe our investigation of extended initial conditions for \emph{single runs}. In particular, we study the agreement of EMT and single-run evolution of the actual system as we vary the spatial extent of the initial condition.

We carry out calculations from exact numerical considerations for systems of 801 sites without changing the configurations of the transition rates once set in accordance with the double-delta distribution, and take initial conditions that are not of the form $P_m(0) = \delta_{m,0}$, but of the extended form
$$P_m(0)= \frac{1}{2 \mu + 1}\sum_{r=-\mu}^{\mu}\delta_{n,r}$$
which represents a \emph{patch} initial condition of spatial extent of $2\mu+1$ sites. We call this the initial width. The limit $\mu=0$ gives us back the initial condition we have used in the studies above. We find that larger patches result in smaller deviations of the EMT predictions from the exact results. 

The inset of Fig. 6 shows values of $D(t)/D(0)$ for a single
configuration
for two different values of the width (open circles represent $\mu =50,$
crosses represent $\mu =5$) along side the corresponding prediction of
effective medium theory (solid line). The integrated difference between
the
EMT result and the exact results depicted in that inset (as plotted on a
logarithmic time scale) provides a convenient measure of the error. We
thus
define, for each value of $\mu ,$ a measure of the relative error,
through
the expression%
\begin{equation*}
E_{R}=\int_{-\infty }^{\infty }\frac{D^{EMT}\left( s\right)
-D^{EX}\left(
t\right) }{D^{EX}\left( s\right) }ds
\end{equation*}%
where $s=\ln \left( \langle f\rangle t\right) .$ A plot of the
(numerically
evaluated) relative error as a function of the initial width $\mu $ is
presented in the main graph in Fig. 6, and clearly shows that the
relative
error decreases monotonically as the patch width increases.

\begin{figure}[]
\centering
\includegraphics{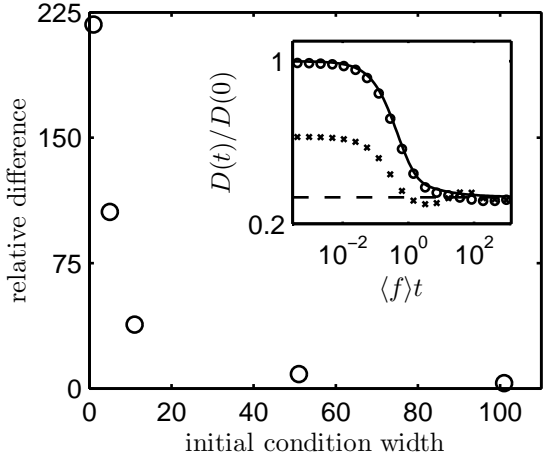}
\caption{Relative difference between exact results and EMT, or error of the EMT, for a single run. Plotted as large open circles is the error (see text for definition)  as a function of the number of sites initially occupied, i.e., the value $2\mu + 1$. No averages are performed. The error is seen to decrease as the initial width increases, allowing the walker to sample different configurations. Inset: Comparison of $D(t)/D(0)$ curves for two different values of the initial width, $\mu=5$ (crosses) and $\mu=50$ (open circles), with the EMT (solid line). It is clearly seen that EMT agrees with the simulations for spatially extended initial conditions without averaging.}
\label{difference_sr_inset}
\end{figure}

\subsubsection{Correlation Type Observables}
There are, in general, different kinds of observables that can be computed
from the solution to the Master equation. Simple observables $O$ are those
which associate with each site (state) $m,$ a value $O_{m}$ that the
observable takes when the particle is  in that state. The mean
value associated with such an observable at any time $t$ can then be written as
\begin{equation}
\langle O\left( t\right) \rangle =\sum_{m}O_{m}P_{m}\left( t\right)
=\sum_{m,n}O_{m}\Psi _{m,n}\left( t\right) P_{n}\left( 0\right),
\end{equation}
where in the second form
we have expressed the result in terms of the propagators $\Psi,$ and the \emph{initial} probability
distribution governing the particle's occupation of the possible states of
the system. This can be put in the form 
\begin{equation}
\langle O\left( t\right) \rangle =\sum_{n}\langle O\left( t\right) \rangle
_{n}P_{n}\left( 0\right)
\end{equation}
where
\begin{equation}
\langle O\left( t\right) \rangle _{n}=\sum_{m}O_{m}\Psi _{m,n}\left(
t\right)
\end{equation}
is the mean value of the observable given that the particle started in state 
$n$ at time $t=0.$ Simple observables can thus be calculated by incorporating into the averaging process an average
over the different possible starting locations of the particle.

Correlation type observables, also of great interest in statistical physics,
do not correspond to (simple) observables of this type. Indeed,
they  \emph{span} two or more different states (or the same state at two different times). An example is $\langle A\left( t\right) B\left(
0\right) \rangle$ given by
\begin{eqnarray}
\sum_{m,n}A_{m}\Psi _{m,n}\left( t\right) B_{n}P_{n}\left(
0\right)
=\sum_{n}\langle A\left( t\right) \rangle _{n}B_{n}P_{n}\left( 0\right). 
\end{eqnarray}
Here $A_{m}$ and $B_{m}$ are, respectively, the values of  $A$ and $B$ when
the particle is in the state $m,$ and
\begin{equation}
\langle A\left( t\right) \rangle _{n}=\sum_{m}A_{m}\Psi _{m,n}\left(
t\right) 
\end{equation}
is the mean value of $A$ at time $t$ if the particle started in state $n$ at 
$t=0$. Consider, for instance  $A^{\ell }$ with components
\begin{equation}
A_{m}^{\ell }=\delta _{m,\ell }.
\end{equation}
It is an indicator observable taking the value $1$ if the particle is at
site $\ell $ and the value $0$ otherwise. Then the correlation function
\begin{equation}
\langle A^{\ell }\left( t\right) A^{\ell ^{\prime }}\left( 0\right) \rangle
=\Psi _{\ell ,\ell ^{\prime }}(t)P_{\ell ^{\prime }}\left( 0\right) 
\end{equation}
is just the propagator $\psi _{\ell ,\ell ^{\prime }}$ weighted by the
relative initial probability of finding the particle in the state $\ell
^{\prime }$. This shows that the propagators $\psi _{\ell ,\ell ^{\prime }}$
themselves can also be considered as observables of the system. Of course,
in a specific disordered system, the self-propagator $\psi _{\ell ,\ell
}\left( t\right) $ , e.g., will depend on the location of site $\ell $ in
the disordered chain. The effective medium propagator $\Psi _{0}\left(
t\right) $ may not, therefore give a good approximation to any given
self-propagator $\psi _{\ell ,\ell }\left( t\right) $ in any single
realization of the disordered system. We intuitively expect, however, that
self-propagators, averaged over an initial distribution of starting
positions on the same chain, will approach that of the effective medium, as
the width of the initial distribution of starting sites is increased, i.e.,
that
\begin{equation*}
\lim_{\mu \rightarrow \infty }\frac{1}{2\mu +1}\sum_{\ell =-\mu }^{\mu }\psi
_{\ell ,\ell }=\Psi _{0}.
\end{equation*}%
To verify this intuition, we present calculations in Fig. 7 of the
self-propagator averaged over such an initial distribution of starting
sites, for different values of $\mu =3,9,15,$ and $25.$ A comparison of the
limiting curve with the predictions of effective medium theory is given in
Fig. 8.
\begin{figure}[]
\centering
\includegraphics{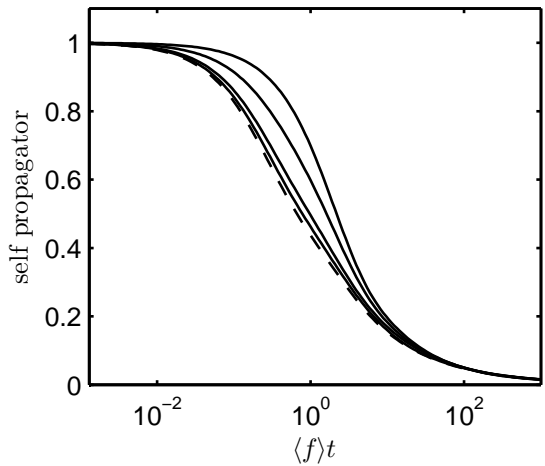}
\caption{Effect of varying patch width of the initial condition. Plotted is the time dependence of the self propagators in a disordered chain (the same disordered chain is used in all of the calculations) whose transfer rates are drawn from a double-delta distribution with $f_1/ f_2 = 0.1$ and $\alpha  = 0.5$. Self propagators at 401 sites, 200 to the left and 200 to the right of origin are calculated. The dashed line is the average of all of the 401 self propagators obtained this way. The solid lines correspond to averaging over 3, 9, 15 and 25 (from top to bottom) of the self propagators around the origin.}
\label{sp_1}
\end{figure}

\begin{figure}[]
\centering
\includegraphics{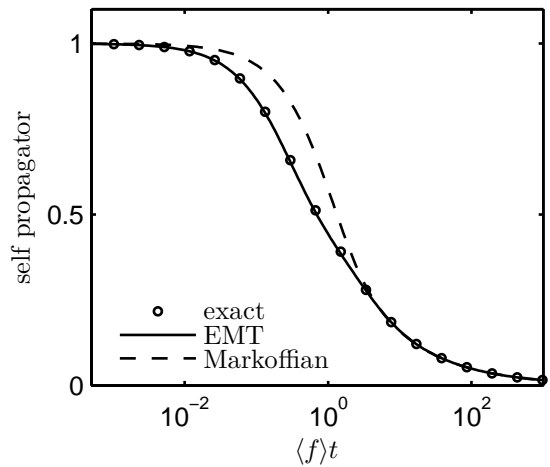}
\caption{Self propagators obtained from the disordered chain (numerically, represented by the open circles), by the effective medium theory (solid line) and from an ordered chain with transfer rates $F_{eff}$ throughout (dashed line).}
\label{sp_2}
\end{figure}

\section{Spatially long range memories}
The original evolution equation that describes the disordered system, Eq. (\ref{1d}), is local in time and nearest-neighbor in the character of its transition rates. Given that effective medium theory provides an approximate rather than exact description of the actual dynamics described by Eq. (\ref{1d}), one may ask whether the introduction of non-locality in time  in the EMT should be accompanied by non-locality in space as well. This is a natural question to pose because of the emergence of spatially long range memories that were found long ago \cite{klongrange,kbook} when GME's were calculated for quantum mechanical systems by a method of Eq. (\ref{prescr}), similar in spirit to the one followed in the EMT. Stated differently, the question we ask is whether the replacement of Eq. (\ref{1d}) by Eq. (\ref{gme2}) with nearest-neighbor transition memories is sufficient or whether the latter should span longer distances. The answer is provided in this section.

Consider Eq. (\ref{1d}) solved for $\tilde{P}_m(\epsilon)$, the Laplace transform of the probability of occupation of the $m$th site in terms of the matrix $A^{\mu}$ corresponding to the configuration $\mu$ (a particular realization of the transition rates $f$ throughout the system). Carrying out the average over the configurations $\mu$ one gets a translationally invariant situation:
\begin{equation}
\tilde{P}_m(\epsilon)=\sum_{n}\left\langle \frac{1}{\epsilon+A^{\mu}} \right\rangle_{m-n} P_n(0).
\end{equation}
Performing a discrete Fourier transform, we get
$\tilde{P}^k(\epsilon)/P^k(0)$ which we substitute in Eq. (\ref{prescr}) to get the \emph{exact} memory function:
\begin{equation}
\tilde{\mathcal{A}}^k=\left\langle \frac{1}{\epsilon+A}\right\rangle^k-\epsilon.
\end{equation}
Because the configuration average has been carried out already at this point, we do not display the superscript $\mu$ on the $A$.

There is no guarantee whatsoever that the $k$-dependence of $\tilde{\mathcal{A}}^k$ is of the form $(1-\cos k)$. The exact memories need not, therefore, have nearest neighbor character. The nature of the disorder will influence the $k$-dependence. It is therefore clear that spatially long range memories will naturally develop, in general, their precise form being determined by the particular distribution $\rho(f)$.

The exact procedure is to be contrasted with the 
EMT procedure, which, as explained in Section 2, necessarily results in the absence of spatially long range memories. This is so because one \emph{assumes} the memories to be nearest-neighbor in character, and obtains them variationally.

\begin{figure}[h]
\centering
\includegraphics{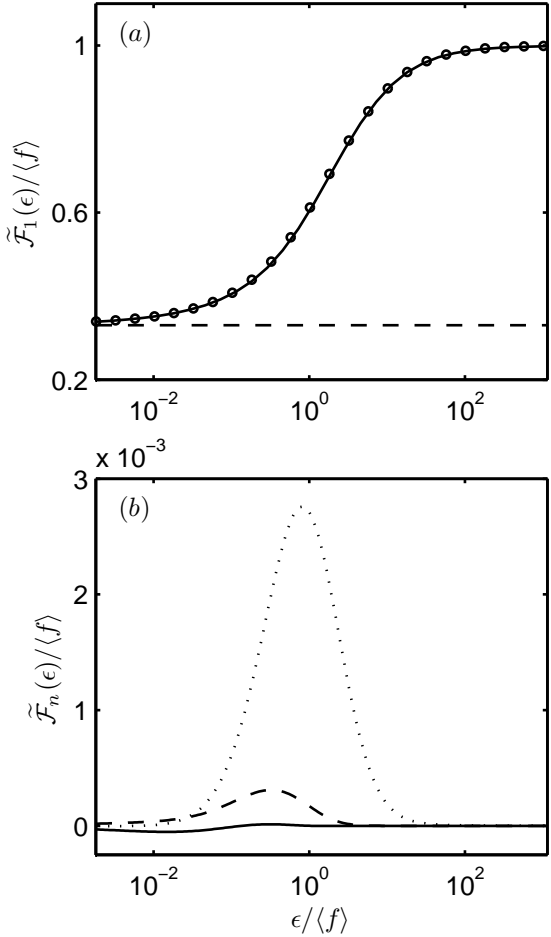}
\caption{Spatially long range memories obtained from exact numerical considerations plotted as a function of the Laplace variable. Units on both axes are of $\langle f \rangle$. Plotted in (a)  is the exact $\widetilde{\mathcal{F}}_1(\epsilon)$ (solid line), calculated for rings of 100 sites, and the almost identical EMT memory (open circles) along with the asymptotic rate $1/\langle 1/f \rangle$ (dashed line). Plotted in (b) on a scale blown up by almost 3 orders of magnitude are the much smaller long range memories $\widetilde{\mathcal{F}}_n(\epsilon)$ in dotted, dashed and solid lines, for $n=2,3,4$, respectively. The distribution is double-delta with $f_1/f_2=0.1$ and $\alpha=0.5$.}
\label{memories}
\end{figure}

In Fig. (\ref{memories}) we display the result of the full numerical exact procedure outlined above carried out on a chain of 801 sites, making sure during each run that  the value of $P_m(t)$ is negligible (comparable to the precision of the machine used) at the boundaries of the chain. 
The distribution used is double-delta, the two rates are in the ratio $f_1/f_2=0.1$ and the concentration of each is equal to the other. We plot in (a) the Laplace transform of the nearest-neighbor memory obtained from the exact procedure (solid line), $\widetilde{\mathcal{F}}_1(\epsilon)$, as a function of the Laplace variable $\epsilon$, both the abscissa and the ordinate being expressed in units of  the average rate $\langle f \rangle$. Also plotted is the result of the EMT procedure (dots) and the dashed line that represents the asymptotic rate $1/\langle 1/f \rangle$. There is hardly any difference in the exact and the EMT result. What this must mean is that the non-nearest neighbor memories must be \emph{much} smaller in magnitude relative to the nearest neighbor $\widetilde{\mathcal{F}}_1(\epsilon)$. This is shown clearly in (b) where the longer range memory transforms, $\widetilde{\mathcal{F}}_n(\epsilon)$, are shown. The scales in the plots in (a) and (b) differ by a little less than 3 orders of magnitude so it is indeed clear that the long-range memories are small. It is thus that the EMT can successfully describe the evolution even though it possesses only nearest-neighbor memories. Note that, while $\widetilde{\mathcal{F}}_1(\epsilon)$ is sigmoidal in shape, the long-range memories seem to peak for intermediate $\epsilon$ and to be negligible for both large and small $\epsilon.$


\section{Finite Size Effects}
To the best of our knowledge, effective medium considerations have been used only on infinitely large systems in the past. We present below useful EMT results for finite rings of $N$ sites, i.e., chains obeying periodic boundary conditions. The self-propagator for such a system is given in the Laplace domain by
\begin{equation}
\tilde{\psi}_0(\epsilon)=\frac{1}{N}\sum_k \frac{1}{\epsilon+2\tilde{\mathcal{F}}
(\epsilon)(1-\cos k)}
\label{sp}
\end{equation}
where $k$ takes on the values $(2\pi/N)[0,1,2,...N-1]$. In the long time limit, the self propagator $\psi_0(t)$ tends to $1/N$ as one knows both from the explicit limit of Eq. (\ref{sp}) or from the physical statement that the probability equalizes over the ring sites. This means via an Abelian theorem that $\epsilon\tilde{\psi}_0(\epsilon) \rightarrow 1/N$ as $\epsilon  \rightarrow 0.$ The use of this limit in Eq. (\ref{genxi}) leads to an important long-time consequence of our general equation (\ref{theeqn}),
\begin{equation}
\frac{1}{F_{eff}}=\frac{N}{N-1}\int_0^{\infty}df\frac{\rho(f)}{f+F_
{eff}(\frac{1}{N-1})},
\label{finitess}
\end{equation}
which is an extension to finite systems of the well-known harmonic mean result of Eq. (\ref{harmo}). Here we have used $F_{eff}=\tilde{\mathcal{F}}(\epsilon \rightarrow 0)$ as earlier.
Equation (\ref{finitess}) must be solved for $F_{eff}$ implicitly and becomes explicit only as $N \rightarrow \infty$ when the $F_{eff}$ term within the integral disappears.

The implicit equation for the case of the double-delta distribution function of Eq. (\ref{doubledelta}),
$$
F_{eff}=\frac{N-1}{N}\left[\frac{\alpha}{f_1+\frac{F_{eff}}{N-1}}+
\frac{1-\alpha}{f_2+\frac{F_{eff}}{N-1}}\right]^{-1},
$$
can be converted into a quadratic equation and solved explicitly. With $$j=f_1(1-N+N\alpha)+f_2(1-N\alpha),$$ one has
\begin{eqnarray}
F_{eff}=\frac{j\pm\sqrt{j^2+4(N-1)f_1f_2}}{2}.
\label{j}
\end{eqnarray}
Normally, i.e., when both $f_1$ and $f_2$ are non-zero, there is a unique solution as we discard the negative root because $F_{eff}$ must be real.

\begin{figure}[]
\centering
\includegraphics[width=3in]{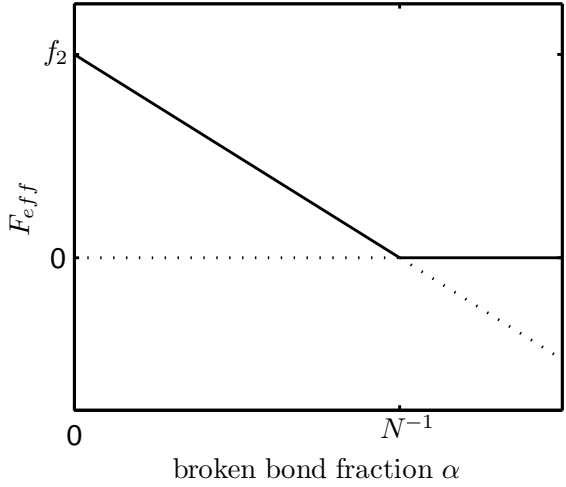}
\caption{Bifurcation of the effective long time transfer rate  for a double delta distribution in a finite system of $N$ sites. Plotted is $F_{eff}$ as  a function of the concentration of broken bonds (i.e., bonds with the rate $f_1=0$), the rate associated with the remaining fraction $1-\alpha$ of unbroken bonds being equal to $f_2$. A transcritical bifurcation occurs when $\alpha$ equals $1/N$. For concentrations higher than this value, the effective rate vanishes but changes linearly with the concentration for lower $\alpha$. Solid (dotted) lines denote the stable (unstable) solution.}
\label{transCB}
\end{figure}

If one of the two possible rates, e.g. $f_1,$ is zero, i.e., if broken bonds exist in the finite system, an interesting situation arises, \emph{both roots} being of physical interest. The lower root is zero, not negative, in this case. If one varies the concentration  $\alpha$ of the broken bonds, a \emph{transcritical bifurcation} occurs as displayed in Fig. \ref{transCB} at the point at which $\alpha$ equals the reciprocal of the number of sites in the ring. As this number increases, the bifurcation point moves towards vanishing concentration. We recover the known result that, for an infinite system, the effective rate is zero for any concentration of broken bonds. Additionally, we get a percolation threshold for finite systems. The two solutions exchange stability at the critical concentration ($\alpha=1/N$), there being transport throughout the ensemble-averaged system for broken bond concentrations below the critical value.

It is interesting to see how the effective medium nearest neighbor memory function compares with the exact nearest neighbor memory function as $\epsilon \to \infty$ in finite rings. One can obtain the exact memory functions for a ring of $N$ sites through Eq. (49) by averaging over all possible configurations of the ring. For simplicity, we will consider the double-delta distribution with $\alpha=1/2$. For rings with $N=2,3,4,$ and $5$ sites we have the exact values, $\lim_{\epsilon\rightarrow0}\widetilde{\mathcal{F}}_1^{EX}(\epsilon)$ 
\begin{align}
&\widetilde{\mathcal{F}}_1^{EX}(0)= \nonumber \\
\text{N=2:\ \ \ \ }  &2f_2\frac{r}{r+1}, \nonumber \\
\text{N=3:\ \ \ \ }  &8f_2\frac{r(r+2)(2r+1)}{(5r+1)(r+5)(r+1)}, \nonumber \\
\text{N=4:\ \ \ \ } &16f_2\frac{r(1+3r)(3+r)(r+1)}{124r(1+r^2) + 230r^2 + 17(1+r^4)}, \nonumber \\
\text{N=5:\ \ \ \ } &16f_2\frac{r(3+2r)(2+3r)(1+4r)(4+r)}{(7+3r)(3+7r)(r+1)(7+36r+7r^2)}. 
\label{memories_exact}
\end{align} 

The effective medium memory function is given by Eq. (\ref{j}) with $\alpha=1/2$.  In order to quantitatively examine how different the exact and effective medium values are, we define a relative difference as
$$\frac{1}{f_2}\left[ \frac{ F_{eff}  - \widetilde{\mathcal{F}}_1^{EX}(0)}{\widetilde{\mathcal{F}}_1^{EX}(0)} \right],
$$
and plot it as a function of $f_1/f_2$ in Fig. 11 for $N=3,4,$ and $5$. We find that the values predicted by the effective medium theory are slightly different from the exact values. The relative difference between the two decreases as the number of sites in the ring becomes larger and larger. Therefore the effective medium theory predicts the correct values in the limit $\epsilon \to \infty$ when  $N \to \infty$, but  finite size effects exist otherwise. Note that for finite $N$, effective medium theory always predicts larger values than those that are calculated exactly.

\begin{figure}[]
\centering
\includegraphics{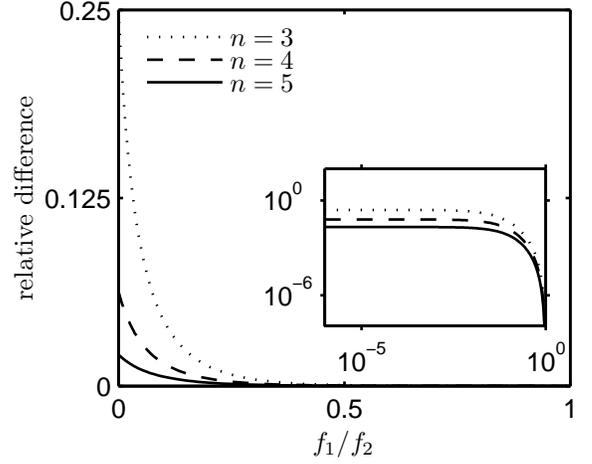}
\caption{The behavior of the relative difference between the nearest neighbor effective rates calculated from the EMT and an exact numerical procedure (see text) as a function of $f_1/f_2$ for a sum of two delta functions with $\alpha = 0.5$}
\label{finite_difference}
\end{figure}

\section{Concluding Remarks}
The purpose of this paper is to make a contribution to the description of transport of quasiparticles such as electrons, excitons, atoms, interstitials, vacancies, or other more formally regarded random walkers in a disordered system such as a solid, a photosynthetic system, or a molecular aggregate. We have focussed our attention on incoherent motion as described by a Master equation and have restricted our analysis to nearest-neighbor transfer on a 1-dimensional chain, infinite or finite obeying periodic boundary conditions (ring). Our general goal is to translate information about static spatial disorder of the given system into dynamic temporal features of a representative ordered system, to do it explicitly by converting distribution functions into memory functions, to study the validity of predictions of the effective medium theory, and to report several extensions we have made of the theory.

Starting with the spirit of effective medium theory used by many \cite{acreview,ancient,other,kirk,amherst,hkk,olax} as expressed through Eq. (\ref{everyone}), we arrive at a transparent prescription Eq. (\ref{theeqn}). The prescription is in the form of a transform, is a natural generalization to all times of well known results such as the harmonic mean recipe of Eq. (\ref{harmo}) for effective long time rates, and involves the solution of  an implicit equation for the Laplace transform of the EMT memory.

We show how to obtain the memory in the time domain by numerical inversions of the Laplace transform produced by the solution of the implicit equation. Additionally we derive, exactly, partial information about the memories. We provide an understanding of the special feature of EMT memories that it consists of two pieces of which one is a delta function. We derive a simple approximate formula in the time domain, Eq. (\ref{expapprox}), for the memory. It can provide a rough and sometimes adequate representation of the exact evolution as Fig. 3 shows for the particular distribution and parameter set that we have used in that case. Lest one develop a false confidence in this coarse approximation, we have shown Fig. 5b in which its predictions for a broken bond system are quantitatively quite different from the exact answers.

We use effective medium theory to go beyond an asymptotic description and compare the EMT description with exact (numerically obtained) predictions. The quantity we choose for comparison is the time-dependent diffusion coefficient $D(t)$ which is proportional to the time derivative of the mean square displacement or equivalently to the time integral of the memory function. Not only do we find excellent agreement at long and short times as expected from previous work, but surprisingly  good agreement at intermediate results also. Figs. 4 and 5 show this clearly. We carry out this comparison in two parts: by doing an ensemble average over initial conditions for  localized initial placement of the walker; and by doing a single-run (no ensemble average over initial conditions) analysis for a spatially extended placement of the walker. The purpose of the latter is to examine the validity of using ensembles for single run situations in patch type initial placement. We also carry out separately configuration averages of the exact and EMT evolutions and find fine agreement. For this latter purpose we choose the selfpropagator as the quantity to calculate.

We also find that in contrast to the EMT treatment, the exact replacement of the disordered system by the ordered system with memory, outlined in the introduction and carried out in detail in section 5, results in spatially long range memories as in earlier analyses of quantum systems \cite{klongrange,kbook}. This is in spite of the fact that the original disordered system has only nearest neighbor rates. We display these memories $\mathcal{F}_n$ which connect a site to another, $n$ sites away. We do this both in the Laplace and the time domains, and find that the nearest neighbor ones are typically larger by an order of magnitude than the others. This explains the success of the EMT even though its memories have only nearest neighbor character.

With the help of our formalism based on Eq. (\ref{theeqn}), we investigate finite size effects on effective medium theory and find interesting new results: corrections depending on size appear in the harmonic mean formula (\ref{harmo}). A novel result emerges involving a bifurcation of the effective long time rate of transfer $F_{eff}$ as the concentration of broken bonds is varied. The bifurcation is transcritical in nature, the vanishing solution for $F_{eff}$ being stable for large concentration of the broken bonds and the nonzero solution being stable for small concentrations relative to a size-dependent critical value.

Thus, we have presented a number of extensions of effective medium theory in this paper. Not discussed here, but important to point out, are other recent extensions along a  line of research recently taken by two of the present authors in their study of transport on small world networks, particularly of the Neumann-Watts kind \cite{parriskenkre,cpk,pck}. In those systems standard rings (finite chains with periodic boundary conditions) with nearest neighbor hopping rates for the random walker form the ordered part and additional small world connections make up the disordered part. Of particular interest to the developments of the present paper is the use of effective medium theory to develop memory functions that connect greater than nearest-neighbor pairs.  Indeed, to correctly describe transport on small world networks, as well as on the partially disordered complex networks of ref. \cite{pck}, it is generally necessary to include memory functions connecting all pairs of sites on the network \emph{except} nearest neighbors, in interesting contrast to what we have shown here to be the case for the 1-d disordered chain.  It is possible that the techniques developed to understand complex networks can be applied to understand the nature of the spatially long range memory functions for disordered systems defined on topologically ordered lattices of the sort we have considered in this paper.

In examining previous work in this field, we find important avenues that were opened by the work of Haus and Kehr \cite{hkk,
kehr,kehr2}. Their approach appears to be similar to ours in spirit. Prescriptions exist in their work for going from disorder to explicit forms of the GME or the CTRW. Furthermore, their use of projection techniques \cite{zw} to the disorder problem includes important considerations of the initial term \cite{zw,kinit} which has been often overlooked \cite{ks} in applications of this technique. We have also carefully examined the question of the usefulness of the widely quoted analysis of ref. \cite{ks}. This question is perhaps important given the absence in that analysis of a practical prescription for obtaining a usable memory function from a quantity describing disorder such as a rate distribution function. Our answer is that  ref. \cite{ks} helped stop the unjustified concerns that some authors \cite{badauthors} seemed to have expressed about  the applicability of the GME, equivalently the CTRW, to disordered systems.  Their message that GME's or CTRW's are fully capable of treating disordered systems is correct and valuable. On the other hand, we have explained in the Introduction to the present paper, how the correctness and applicability of GME/CTRW's can be understood without the need for detailed argument. The real need is a practical prescription for the translation of disorder features into the time dependence of memories or pausing time distributions. The development and use of such a prescription, already apparent in early work \cite{sl,kehr,kehr2}, has been attempted in the present paper.

In concluding, for the use of those who prefer to work with CTRW's rather than GME's, we give explicit expressions for the CTRW pausing time distribution functions in effective medium theory. The situation here (in the EMT, not in the exact system as can been seen in Fig. \ref{memories}) is separable in time and space, and so could be addressed by the formula in ref. \cite{kms}. However we exploit the \emph{general} relation given first by Kenkre and Knox \cite{kk,K} in Eq. (43) of the first of those references or Eq. (30) of the second. Corresponding to the GME, Eq. (\ref{gme2}),  the CTRW equation which is a sort of non-Markoffian Chapman-Kolmogorov equation, becomes \cite{K} 
\begin{equation}
P_m(t)=P_m(0)\left[1-\int_0^t ds \Psi(s)\right]+\int_0^t ds \sum_n\mathcal{Q}_{mn}(t-s)P_n(s)],
\label{ctrw}
\end{equation}
with
\begin{eqnarray}
\tilde{\mathcal{Q}}_{mn}(\epsilon)&=&(\delta_{m,n+1}+\delta_{m,n-1})\left[\frac{\tilde{\mathcal{F}}(\epsilon)}{\epsilon+2\tilde{\mathcal{F}}(\epsilon)}\right], \nonumber \\
\tilde{\Psi}(\epsilon)&=&\frac{2\tilde{\mathcal{F}}(\epsilon)}{\epsilon+2\tilde{\mathcal{F}}(\epsilon)}.
\label{pausing}
\end{eqnarray}
These formulae can be used after the determination of the EMT memory in the various ways we have explained.

This work was supported in part by the NSF under grant no. INT-0336343 and by the Program in Interdisciplinary Biological and Biomedical Sciences at UNM 
funded by the Howard Hughes Medical Institute. One of the authors (VMK) thanks D. Tchawa for insightful remarks and his dogged determination in discussions.


\begin{thebibliography}{99}
\bibitem{feetnote}
If $x$ is a random variable,  $\left\langle e^{-\lambda x}\right\rangle
=e^{-\lambda \left\langle x\right\rangle }\left\langle e^{-\lambda \left( x-\left\langle
x\right\rangle
\right) }\right\rangle .$ Using $y=-\lambda(x-\langle x \rangle)$ in the inequality $e^{y}\geq 1+y,$  one gets $$
\left\langle e^{-\lambda x}\right\rangle \geq e^{-\lambda \left\langle x\right\rangle }\left\langle
1-\lambda \left( x-\left\langle x\right\rangle \right) \right\rangle =e^{-\lambda \left\langle
x\right\rangle }
$$
which reduces to the well-known result $
\left\langle e^{-\lambda x}\right\rangle \geq e^{-\lambda \left\langle x\right\rangle }.
$
The integration
$$
\int_{0}^{\infty }d\lambda \;\left\langle e^{-\lambda x}\right\rangle =\left\langle
\int_{0}^{\infty }du\;\frac{e^{-u}}{x}\right\rangle =\left\langle
\frac{1}{x}\right\rangle
\geq \int_{0}^{\infty }d\lambda \;e^{-\lambda \left\langle x\right\rangle
}=\frac{1}{%
\left\langle x\right\rangle }
$$
yields
$
\left\langle \frac{1}{x}\right\rangle \geq \frac{1}{\left\langle x\right\rangle }
$
and hence $
\left\langle x\right\rangle \geq \left\langle \frac{1}{x}\right\rangle ^{-1}.
$
\end{thebibliography}
\end{document}